\documentclass[prl, twocolumn, amsmath, amssymb, superscriptaddress, showpacs]{revtex4}
\usepackage{graphicx}
\usepackage{epsfig}
\usepackage{epstopdf}

\begin{document}
\title{Magnetoresistance measurements of Graphene at the Charge Neutrality Point}

\author{Yue Zhao}
\affiliation{Department of Physics, Columbia University, New York, NY 10027}

\author{Paul Cadden-Zimansky}
\affiliation{Department of Physics, Columbia University, New York, NY 10027}

\author{Fereshte Ghahari}
\affiliation{Department of Physics, Columbia University, New York, NY 10027}

\author{Philip Kim}
\affiliation{Department of Physics, Columbia University, New York, NY 10027}

\date{\today}

\begin{abstract}

We report on transport measurements of the insulating state that forms at the charge neutrality point of graphene in a magnetic field. Using both conventional two-terminal measurements, sensitive to bulk and edge conductance, and Corbino measurements, sensitive only to the bulk conductance, we observed a vanishing conductance with increasing magnetic fields. By examining the resistance changes of this insulating state with varying perpendicular and in-plane fields, we probe the spin-active components of the excitations in total fields of up to 45 Tesla. Our results indicate that $\nu=0$ quantum Hall state in single layer graphene is not spin polarized.

\end{abstract}
 \maketitle

Under a magnetic field, the linear dispersion relation of low energy electron spectrum in graphene leads to unique Landau levels (LLs) whose energy difference is unequally spaced~\cite{Zheng, Gusynin, Peres}. The LL spectrum, given by $E_n=\pm\sqrt{2n\hbar{v_F}^2eB/c}$, where $v_F$ is the Fermi velocity and $n=0, \pm 1, \pm 2, ...$ is LL index, contains an $n=0$ level, termed the zero-energy LL (ZLL). In the absence of appreciable interactions or Zeeman splitting, each LL has a 4-fold degeneracy arising from a real spin and valley degeneracy. The appearance of the quantum Hall (QH) effect in graphene at the LL filling fractions $\nu=\pm2, \pm6,...$ is a manifestation of this 4-fold degeneracy of graphene LLs~\cite{Novoselov, Zhang}. In the high magnetic field regime, however, this effective SU(4) spin-pseudospin symmetry can be broken, with more QH plateaus appearing at $\nu=0, \pm 1, \pm 4$ and developing signatures of QH states for other integer filling fractions~\cite{Zhang1, Jiang, Kurganova}. The $\nu=0$ filling factor that appears at the center of the ZLL presents something of a paradox in QH physics, as it is not marked by the usual longitudinal resistance minima that typify all other filling factors. While initial measurements on disordered samples at this filling factor reported high-field (above 30~T) resistance in the regime of tens of K$\Omega$s~\cite{Abanin}, subsequent reports on this quantum Hall state have shown a strong insulating behavior as sample mobility is increased~\cite{Checkelsky, Checkelsky1, liyuan, liyuan1}, with two-terminal measurements of the highest mobility suspended samples measuring into the G$\Omega$ range at fields as low as 5 Tesla~\cite{Du, Bolotin}.

Theoretically, various models of symmetry breaking and ordering underlying this $\nu=0$ insulating state have been proposed. Most of the models fall under the framework of exchange-driven quantum Hall ferromagnetism that separates different sectors of the SU(4) spin-pseudospin space~\cite{Nomura, Alicea}. These include: a fully spin-polarized ferromagnet~\cite{Abanin2, Shimshoni}, a fully pseudospin-polarized charge density wave~\cite{Jung, Herbut}, a Kekule distortion with a spontaneous ordering of pseudospin~\cite{Fuchs, Nomura2, Hou}, and a canted antiferromagnet~\cite{Kharitonov}. An alternative approach is based on magnetic catalysis: long-range electron-electron interactions that induce an excitonic gap~\cite{Gusynin2}. Experimental reports on the nonzero filling fraction~\cite{Zhang1, Jiang} suggest that the excitations of the $\nu=1$ state have no spin, while the Kosterlitz-Thouless insulating behavior of $\nu=0$~\cite{Checkelsky1} is consistent with a Kekule distortion origin. The various models of the broken symmetry states involve unique bulk spin/pseudospin textures and corresponding edge state configurations~\cite{Abanin2, Fertig}. Thus transport measurements require careful comparison of the bulk and edge state conduction in order to answer questions related to the nature of the symmetry breaking at $\nu=0$.

\begin{figure}[]
    \centering
    \includegraphics[width=1.0\linewidth]{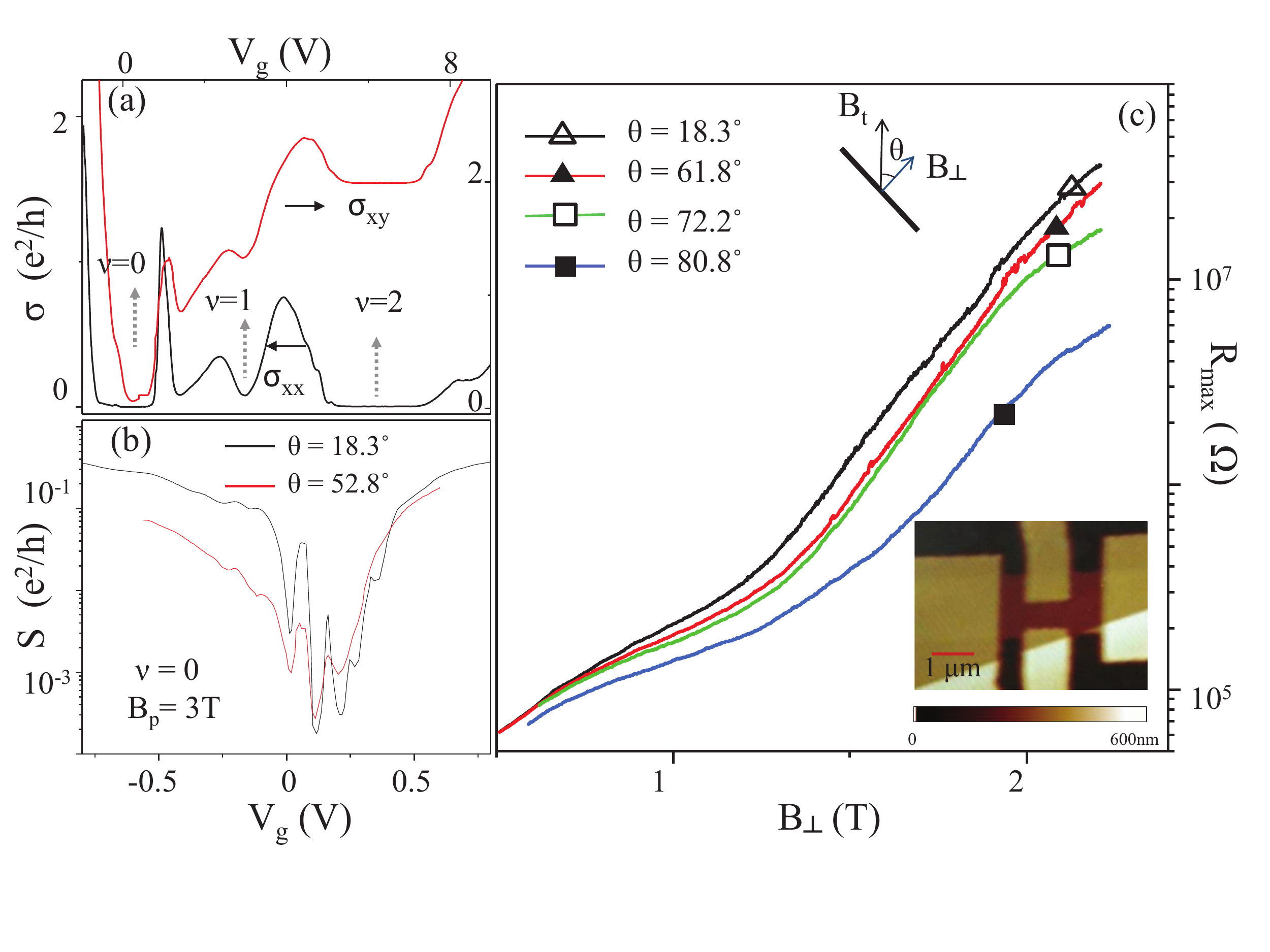}
    \caption{(a) Longitudinal conductivity($\sigma_{xx}$ in black) and hall conductivity($\sigma_{xy}$ in red) as a function of back gate voltage at $T=1.7$~K with $B=4$~T; dashed arrows indicate QH states $\nu=$~0, 1, and 2; (b) Conductance ($S$) on a logarithmic scale versus $V_g$ at $B_\perp=3$~T for two different tilting angles: $\theta=$18.3$^\circ$ and $\theta=$52.8$^\circ$; (c) Maximum two-terminal resistance as $B_\perp$ increases for different tilting angles: $\theta=$18.3$^\circ$ (open triangle), 61.8$^\circ$ (solid triangle), 72.2$^\circ$(open square), and 80.8$^\circ$(solid square); the experimental error of $\theta$ is 0.3$^\circ$; lower right inset: AFM image of the measured suspended graphene device; upper inset shows a schematic diagram for tilting angle and perpendicular and total magnetic field components.}
    \label{Fig.1}
\end{figure}

In this letter we investigate the spin response of the $\nu=0$ QH state in monolayer graphene by measuring the bulk and edge conduction as a function of in-plane magnetic field using high-mobility suspended graphene and on-substrate graphene Corbino device. Our experiments reveal a vanishing conductance at $\nu=0$, but neither exhibits an increasing gap with increasing in-plane field, suggesting that the $\nu=0$ state is not spin-polarized.

The suspended graphene devices are prepared using the methods described in reference~\cite{Bolotin1}: after thermally evaporating Cr/Au electrical contacts onto the mechanically exfoliated graphene samples\cite{Novoselov1}, a chemical etch of buffered hydrofluoric acid is performed to remove the SiO$_2$ under the graphene sample, leaving the whole device suspended approximately 200~nm above the SiO$_2$/Si substrate. An atomic force microscope (AFM) image of the device is shown in the inset of Fig.~1(c). DC current annealing is then performed at low temperature ($T=1.7$~K) to remove residual impurities from the suspended graphene. Four-terminal transport measurements are conducted using conventional low-frequency lock-in techniques. The carrier density of the graphene is tuned by applying back gate voltage $V_g$ to the degenerately-doped Si substrate, with the magnitude of the tuned density determined using Hall measurements. The mobility of this annealed device is $\sim$80,000~cm$^2$/V$\cdot$s. In Fig.~1(a), we show the longitudinal conductivity $\sigma_{xx}$ and Hall conductivity $\sigma_{xy}$ versus back gate at $B=$4~T normal to the graphene basal plane.  As indicated by the vertical arrows, along with clearly developed $\nu=2$ QH state, strong $\nu=0$ and developing $\nu=1$ are observed as plateaus in $\sigma_{xy}$ and the suppression of $\sigma_{xx}$ at the corresponding filling fractions. The appearance of the $\nu=0$ and $\nu=\pm1$ QH states indicates that the four-fold degeneracy of the ZLL is completely broken.

To discern whether the $\nu=0$ symmetry breaking is spin-active, we apply a sequence of tilted magnetic fields that fix the perpendicular magnetic field $B_\perp$ while varying the total magnetic field $B_t$. By fixing $B_\perp$ the magnetic length $l_B=\sqrt{\hbar/eB_\perp}$, Coulomb energy scale $E_{e-e}=e^2/4\pi\epsilon_0\epsilon_r l_B$ are held constant, meaning the electron-electron and exchange interactions that underlie the $\nu=0$ state are unchanged.  However, if this state is fully spin polarized, the current-carrying excitations will have net spins that will be affected by changes in $B_t$ via the Zeeman energy $\Delta E_z=g\mu_B B_t$, where $g$ is g-factor of electron and $\mu_B$ is the Bohr magneton.  At a fixed temperature the changes to the carrier excitation energy will result in a change in conductance observed at the $\nu=0$ filling factor.  Thus by tuning only the Zeeman energy and examining changes in the conductance, we can determine if the activation of the $\nu=0$ state is spin-sensitive.

The results of measuring the insulating state of the suspended device at several different tilting angles are shown in Fig. 1(c), where the resistance maximum $R_{max}$, is measured at the charge neutrality point $V_g=V_D$, at a fixed base temperature $T=$1.6~K. Since the resistance for $\nu=0$ QH state tends to increase rapidly as a function of $B$ in ~\cite{Checkelsky, Checkelsky1, Du, Bolotin}, $R_{max}$ is a good measure to probe this insulating state. Here we use two-terminal current measurement with a constant voltage bias in order to eliminate any self-heating effects ($\leq$~pW) and to maximize the measurable resistance range. At $T=1.7$~K, we found that $R_{max}$ increases from $\sim$10~K$\Omega$ up to 100~M$\Omega$ (comparable to the limit of our measurement set-up) as $B_\perp$ changes from 0 to 3~T. The tilting angle dependence of $R_{max}$ versus $B_\perp$ curves show such a trend: while we do not observe appreciable dependence of $R_{max}$ on in-plane magnetic field at lower values of the tilting angle $\theta$ (i.e., larger $B_\perp/B_t$ ratio), there is an indication that $R_{max}$ decreases at larger $\theta$ (i.e, smaller $B_\perp/B_t$ ratio). This trend becomes most obvious for the largest tilting angle we measured, $\theta=$80.8$^\circ$, corresponding to $B_\perp/B_t$, where we observe that $R_{max}$ versus $B_\perp$ curve is substantially lower than any other curves in the graph.  The observed trend in the suspended device, i.e., decreasing $R_{max}$ with decreasing $B_\perp/B_t$ at fixed $B_\perp$ suggests that the $\nu=0$ gap decreases as $B_t$ increases. This dependence can be viewed as strong evidence against a fully spin-polarized ordering of the $\nu=0$ QH state, as this ordering would result in an increase in the gap as $B_t$ increases. The relative insensitivity of $R_{max}$ to changes in angle for the small tilt angles may be due to broadening induced by thermal smearing or disorders.

There are two obstacles in using suspended samples to draw more quantitative conclusions about the nature of the $\nu=0$ QH state.  First, due to the mechanical instability of suspended samples, $R_{max}$ drifts slightly with respect to $V_g$. Fig.~1(b) shows the conductance as a function of $V_g$ measured at two different tilting angles. Although the overall behavior is consistent, the position of $V_g$ where $R_{max}$ occurs is slightly shifted. Even worse, this shift changes when the device is thermally cycled, making it difficult to estimate the energy gap by the thermally-activated behavior. Second, the four-/two-terminal device geometry measures both the bulk conductance and any possible edge conductance in parallel. This becomes a major source of ambiguity in distinguishing whether the observed insulating behavior originates from the bulk insulating state without the edge conduction or from the localization of edge states by spin/pseudospin-flip scattering~\cite{Abanin, Abanin2}. In order to avoid the mechanical instability and to isolate the bulk conductance, we employ an on-substrate Corbino geometry, a disk-shaped sample with coaxial contacts in which the current flows radially from an inner contact to an outer ring contact.  This geometry not only eliminates any unknown edge effects that might interfere with determining the $\nu=0$ conductance, but is also insensitive to the formation of the known quantized edge conductances of other filling factors. This geometry then directly allows probing bulk conduction, and thus puts the $\nu=0$ insulator on an even footing with the bulk insulating character of every other filling factor~\cite{Das Sarma1}.

\begin{figure}[]
    \centering
    \includegraphics[width=1.0\linewidth]{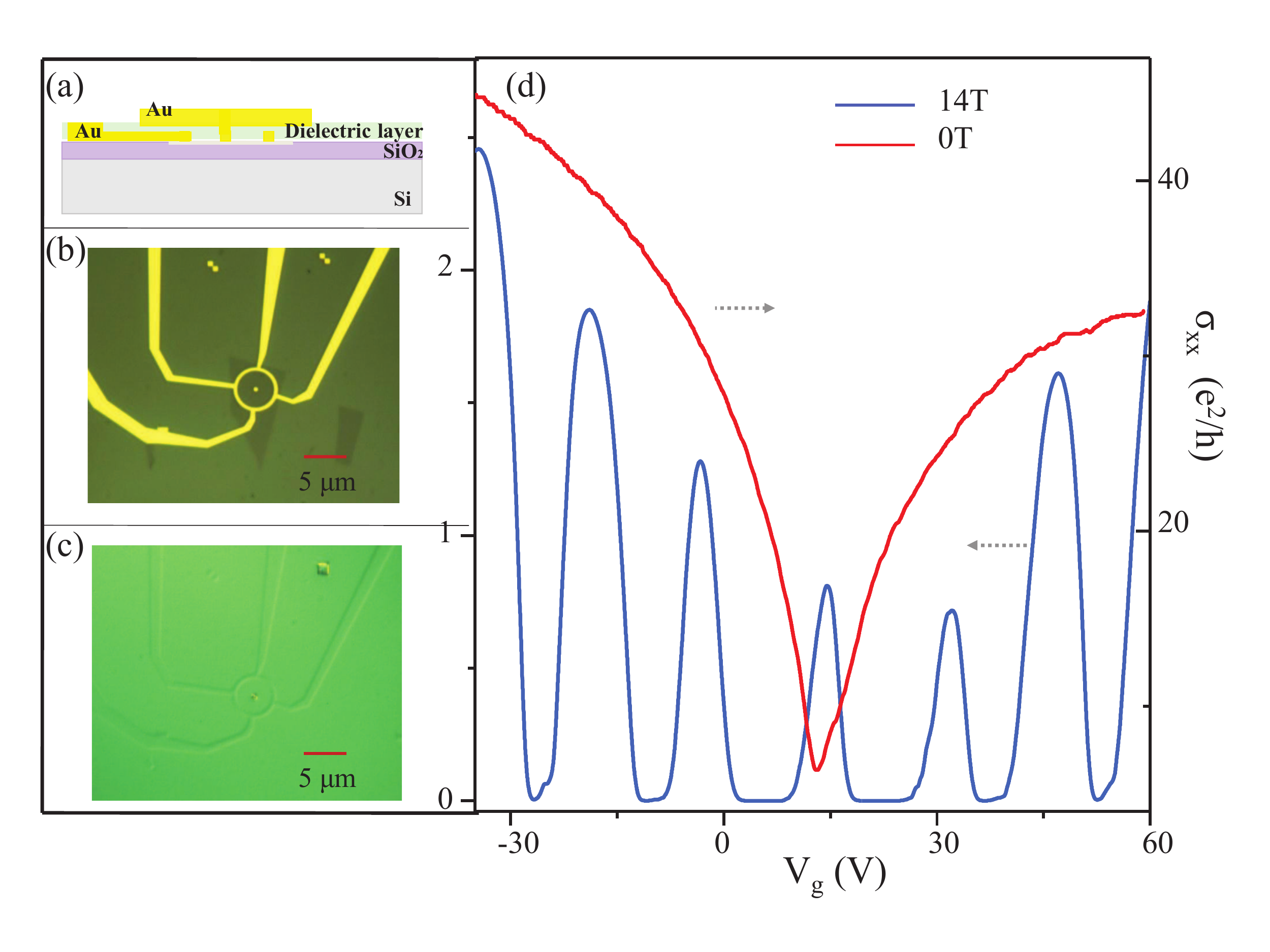}
    \caption{(a) Schematic side view of a Corbino device; (b) Optical device image before the inner electrode is contacted by the ground plate; (c) Optical image of a finished device; (d) Bulk Conductivity $\sigma_{xx}$ as a function of back gate voltage $V_g$ at zero field (in red) and 14~T (in blue), at $T=$7~K.}
    \label{Fig.1}
\end{figure}

The fabrication procedure for our Corbino devices is shown as in Fig.~2(a). Monolayer graphene pieces are deposited on SiO$_2$(300nm)/Si substrates using established mechanical exfoliation techniques, then Au/Cr ring-like electrodes are fabricated by e-beam lithography, (an optical image is shown in Fig.~2(b)), followed by a dielectric layer deposition and a top Au/Cr plate contact to connect to the inner contact (as shown in Fig.~2(c)). The plate geometry connecting to the inner contact guarantees that any voltage applied to this contact will result in a uniform change to the graphene carrier density. To measure the bulk conductance of the graphene, we apply an AC voltage bias ($V_{bias}$) across the inner and outer contacts, and measure the current ($I$) using a current preamplifier and lock-in amplifier. The bulk conductivity is then given by $\sigma_{xx}=(\ln(r_{out}/r_{in})/2\pi)(I/V_{bias})$, where $r_{out}$ and $r_{in}$ are the radii of the outer and inner contacts, respectively.

Changing the back gate voltage $V_g$, we can tune the carrier density in the graphene channel connecting the inner and outer contacts of the Corbino device. Fig.~2(d) shows the bulk conductivity $\sigma_{xx}$ vs. back gate voltage $V_g$, at $B=$0~T and 14~T at temperatures lower than 7~K. The mobility of this particular Corbino device is $\sim$13,000 cm$^2$/V$\cdot$s, obtained from the zero-field resistance. At $B=14$~T, the four-fold degenerate QH state filling factors $\nu=\pm 2, \pm 6, \pm 10$ appear as vanishing $\sigma_{xx}$ at their corresponding carrier density. The gate capacitance of this device is estimated to be $C_g/e=7.1\times10^{10}$cm$^{-2}$V$^{-1}$ from the position of the observed conductivity minima.

 \begin{figure}[]
    \centering
    \includegraphics[width=1.0\linewidth]{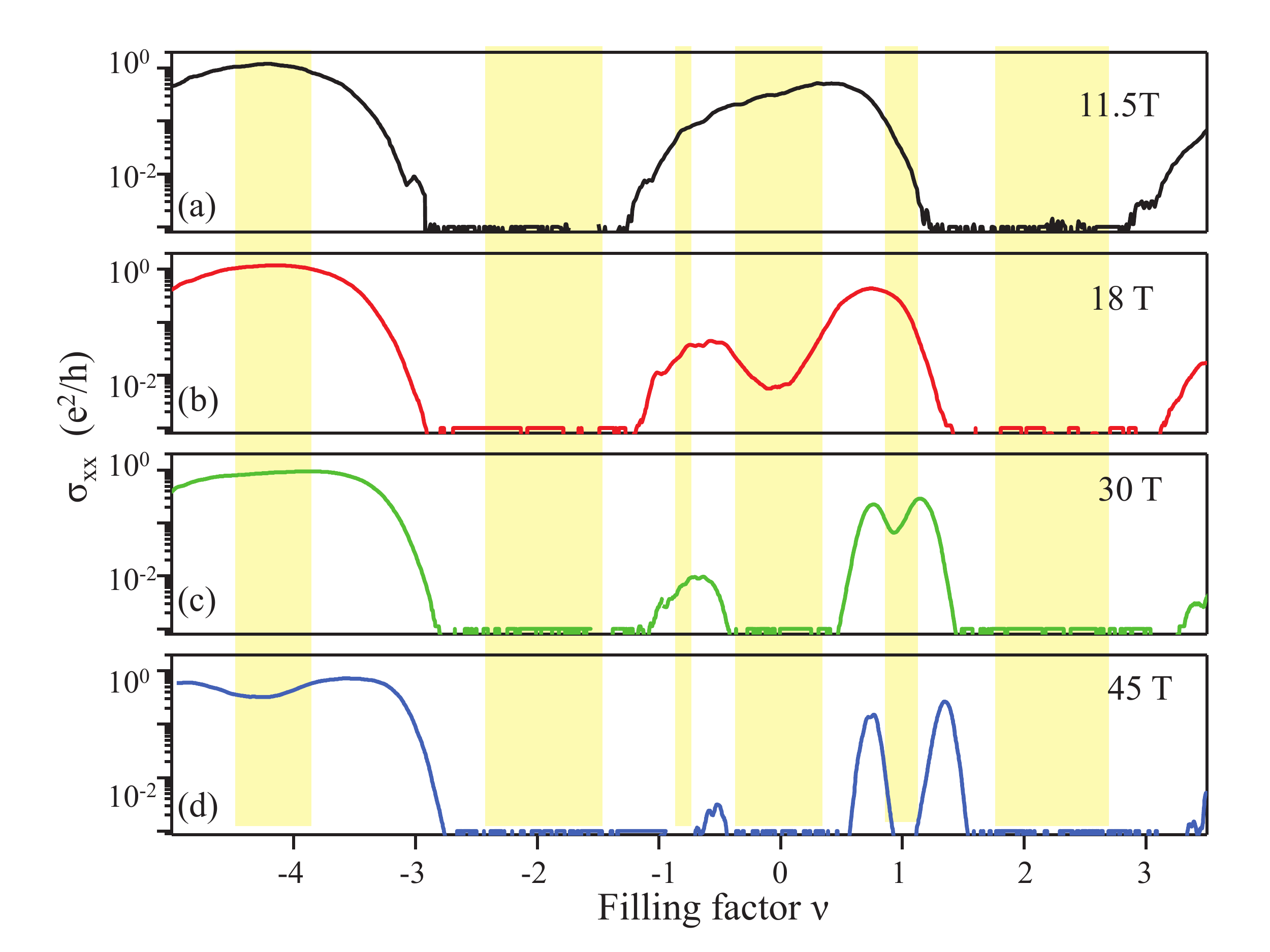}
    \caption{(a) $\sigma_{xx}$ as a function of filling factor at four perpendicular magnetic fields at $T=1.7$~K. The shaded bands highlight the developing filling factors as the four-fold degeneracy of the ZLL is broken, each manifested as a vanishing bulk conductance.}
    \label{Fig.3}
\end{figure}

 Since the mobility of the on-substrate Corbino devices is lower than that of the suspended devices, relatively higher magnetic fields are required to access the degenerately broken filling factors. As shown in Fig.~3, at low field ($B=11.5$~T) well-defined $\nu=\pm2$ states are observed on both sides of the charge neutrality point, indicative of the four-fold QH degeneracy. As the magnetic field increases to 18~T, a dip of bulk conductivity appears at the charge neutrality point. This dip fully evolves and the current flow falls below the noise level at $B=30$~T. This observation of a vanishing bulk conductivity is consistent with the formation of the $\nu=0$ QH state~\cite{Das Sarma1}. At the same magnetic field, the conductivity minima corresponding to the $\nu=\pm 1$ filling factors are visible. At $B=45$~T, the four-fold degeneracy at the zero energy level is completely lifted, and the LL splitting at $\nu=-4$ that marks the degeneracy breaking of the $n=1$ LL is apparent, similar to the previous observation~\cite{Zhang1}. In all measured devices, the magneto-conductance is strongly suppressed in the regime between the $\nu=-1$ and $\nu=-2$ filling factors and is not measurable within our experimental sensitivity, which remains not fully understood.

As with the suspended devices, we adjust the relative strengths of the Zeeman and Coulomb energy in the Corbino devices by tilting the field in order to explore the nature of the $\nu=0$ degeneracy breaking. In Fig.~4(a), $\sigma_{xx}$ vs. filling factor $\nu$ is plotted with constant normal field ($B_\perp=21$~T) and with the total field ($B_t$=$B_\perp/\cos\theta$) increasing. Taking the dielectric constant $\epsilon_r$=4, the characteristic Coulomb interaction energy at $B_\perp=21$~T is $E_{e-e}=740$~K, while the Zeeman energy varies from $E_z=47$~K at $B_t=35$~T to $E_z=60$~K at $B_t=45$~T. As the Zeeman energy is increased, the behavior of the $\nu=0$ and $\nu=4$ states are completely different. For the $\nu=\pm4$ QH state, the $\sigma_{xx}$ minima decrease with increasing $B_t$, indicating that a spin polarization underlies this LL, a finding consistent with previous experiments on Hall bar devices~\cite{Zhang1}. In contrast, the conductance curves of the $\nu=0$ state coincide with each other as the total field is increased from $B_t=35$~T to 45~T. The conductance minima are unvarying even in a magnified logarithmic-scale view, as shown in the middle inset of Fig.~4(a). The fact that this minima is independent, within disorder broadening, to changes in the in-plane field is also consistent with a state that is not fully spin-polarized, and adds further credence to the hypothesis that the $\nu=0$ symmetry breaking is not of spin origin.

We also perform fine-tuned tilted field measurements in a range where the change of Zeeman energy is larger (increased by 50\%) and the $\nu=0$ minima are more sensitive to small changes in $B_\perp$. Fig.~4(b) shows a log-scale $\sigma_{xx}$ vs. filling factor $\nu$ at $B_\perp=14$~T and 15~T. As the normal field increases by $\sim 6\%$, there is a decrease in the bulk conductivity minima, showing that the $\nu=0$ state is not yet fully developed. Increasing the total field by $\sim 50\%$ while fixing $B_\perp$, the minima display the same insensitivity to in-plane field as in Fig.~4(a), reaffirming that the excitations of the $\nu=0$ state have no net spin. We are aware of not observing a decreasing $R_{max}$ with decreasing $B_\perp/B_t$ at fixed $B_\perp$ in Corbino device, although the range of the Zeeman to Coulomb energy ratio change in Corbino device is similar to that of the suspended device. The discrepancy of the behaviors could be understood to be consequences of the different disorder energy scale.

\begin{figure}[]
    \centering
    \includegraphics[width=1.0\linewidth]{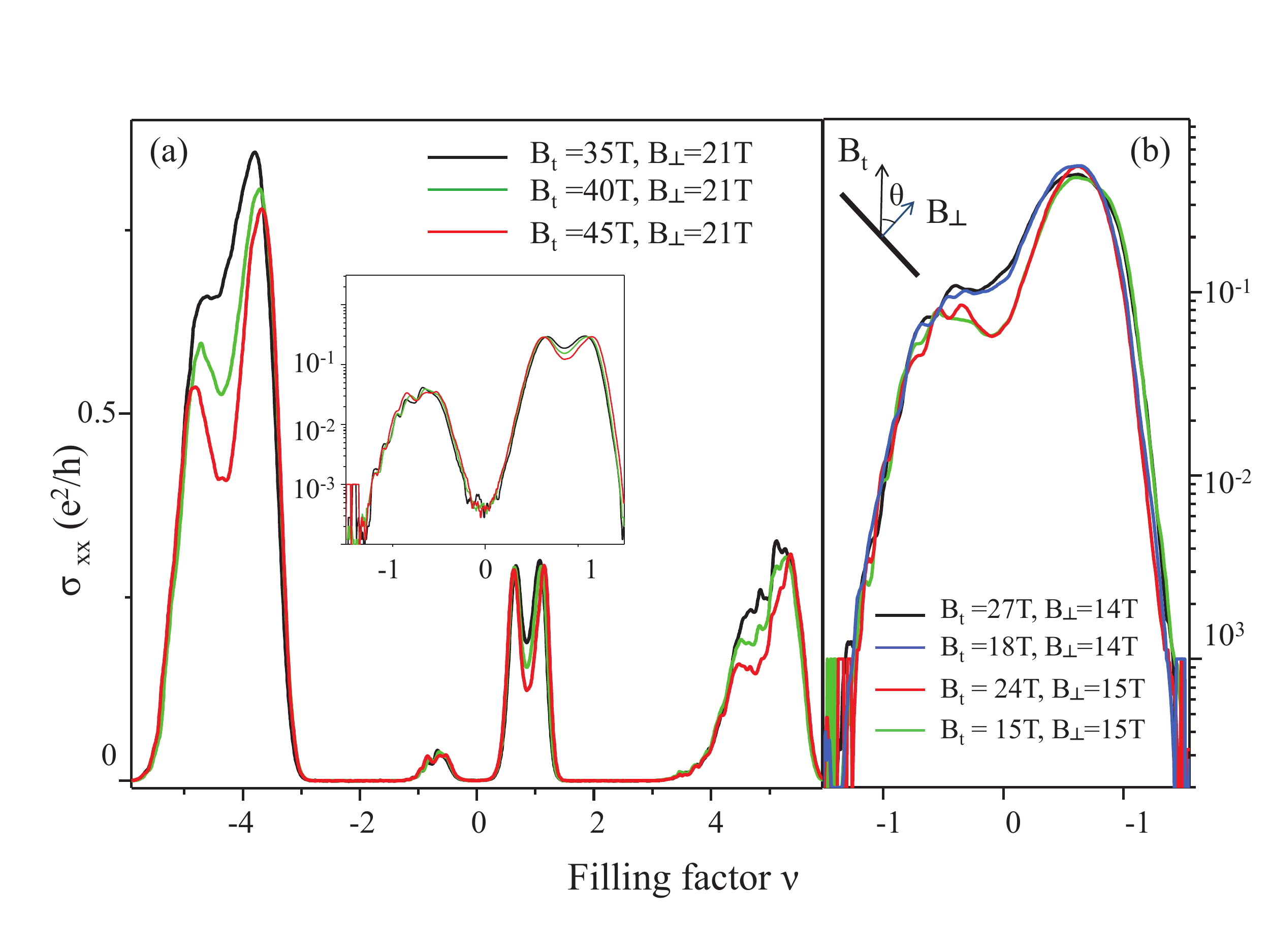}
    \caption{(a) $\sigma_{xx}$ as a function of filling factor at a constant perpendicular magnetic field $B_\perp=21$~T for several different total magnetic field. Data taken at $T=1.7$~K. From top to bottom the corresponding total fields are $B_t=35$, 40, and 45~T, respectively. The inset shows a close-up of the main panel at the $\nu=0$ QH state. (b) $\sigma_{xx}$ at $B_\perp=14$ and 15~T, where it is sensitive to small changes in the perpendicular field.}
    \label{Fig.4}
\end{figure}

As to the $\nu=1$ QH state, experimental data of its tilted-field dependence is also shown in Fig.~4(a). The $\sigma_{xx}$ minima at this filling factor decrease as $B_t$ increases. This observation implies that the origin of this state is due in part to a lifting of real spin degeneracy. Combining this observation with that of the $\nu=0$ state, it produces a symmetry-breaking picture of the ZLL where a non-spin polarized state forms at $\nu=0$ and a spin-polarized state with spin-flip excitations forms at $\nu=1$~\cite{Young2011}.

We are also aware that the observation of the spin-active $\nu=1$ character is inconsistent with the observations of Jiang et al.~\cite{Jiang}, whose measurements implied that excitations of $\nu=\pm 1$ QH states has not spin flip. This raises the possibility that the excitations at $\nu=1$ and its ground state may depend on the specific disorder concentration in individual samples~\cite{Nomura, Goerbig}. However, the fact that the insulating $\nu=0$ state does not respond to increasing in-plane magnetic fields in both suspended and Corbino devices, where disorder densities are very different, provides evidence that disorder effects do not alter our conclusion that the $\nu=0$ QH state is not a spin-polarized state for a wide range of disorder.

The authors thank I. Aleiner, Y. Barlas, E.A. Henriksen, Z. Jiang and A.F. Young for helpful discussions, and thank S.T. Hannahs, E.C. Palm, and T. P. Murphy for their experimental assistance. This work is supported by DOE (No. DEFG02-05ER46215). A portion of this work was performed at the National High Magnetic Field Laboratory, which is supported by NSF Cooperative Agreement, by the State of Florida, and by the DOE (No. DMR-0654118).


\begin{thebibliography}{text}
\bibitem{Zheng}Y. Zheng and T. Ando, Phys. Rev. B {\bf 65}, 245420 (2002).
\bibitem{Gusynin}V.P. Gusynin, S.G. Sharapov, Phys. Rev. Lett. {\bf 95}, 146801 (2005)
\bibitem{Peres} N.M.R. Peres, F. Guinea, and A.H. Castro Neto, Phys. Rev. B {\bf 73}, 125411 (2006).
\bibitem{Novoselov}K.S. Novoselov, et al., Nature {\bf 438}, 197 (2005).
\bibitem{Novoselov1} K. S. Novoselov et al., Science {\bf 306}, 666?-669 (2004).
\bibitem{Zhang}Y. Zhang, Y.W. Tan, H.L. Stormer, and P. Kim, Nature {\bf 438}, 201 (2005).
\bibitem{Zhang1} Y. Zhang et al., Phys. Rev. Lett. {\bf 96}, 136806 (2006).
\bibitem{Jiang} Z. Jiang, Y. Zhang, H. L. Stormer, and P. Kim, Phys. Rev. Lett. {\bf 99},106802 (2007).
\bibitem{Kurganova} E. V. Kurganova, et. al. Phys. Rev. B {\bf 84}, 121407 (2011).
\bibitem{Checkelsky} J. G. Checkelsky, L. Li, and N. P. Ong, Phys. Rev. Lett. {\bf 100}, 206801 (2008).
\bibitem{Checkelsky1} J. G. Checkelsky, L. Li, and N. P. Ong, Phys. Rev. B. {\bf 79}, 115434 (2009).
\bibitem{liyuan} L. Zhang, et al., Phys. Rev. B. {\bf 80}, 241412R, (2009).
\bibitem{liyuan1} L. Zhang, Y. Zhang, M. Khodas, T. Valla, and I. A. Zaliznyak, Phys. Rev. Lett. {\bf 105}, 046804, (2010)
\bibitem{Du} X. Du, I. Skachko, F. Duerr, A. Luican, and E. Y. Andrei, Nature {\bf 462}, 192 (2009).
\bibitem{Bolotin} K. I. Bolotin, F. Ghahari, M. D. Shulman, H. L. Stormer, and P.Kim, Nature {\bf 462}, 196 (2009).
\bibitem{Nomura} K. Nomura and A. H. MacDonald, Phys. Rev. Lett. {\bf 96}, 256602 (2006).
\bibitem{Alicea}J. Alicea and M. P. A. Fisher, Phys. Rev. B {\bf 74}, 075422 (2006).
\bibitem{Goerbig} M. O. Goerbig, R. Moessner, and B. Doucot, Phys. Rev. B {\bf 74}, 161407 (2006).
\bibitem{Abanin} D. A. Abanin et al., Phys. Rev. Lett. {\bf 98}, 196806 (2007).
\bibitem{Abanin2} D. A. Abanin, P. A. Lee, and L. S. Levitov, Phys. Rev. Lett. {\bf 96}, 176803 (2006).
\bibitem{Fertig} H. A. Fertig and L. Brey, Phys. Rev. Lett. {\bf 97}, 116805 (2006).
\bibitem{Herbut} I. F. Herbut, Phys. Rev. B {\bf 75}, 165411 (2007).
\bibitem{Shimshoni} E. Shimshoni, H. A. Fertig, G. V. Pai, Phys. Rev. Lett. {\bf 102}, 206408 (2009).
\bibitem{Jung} J. Jung and A. H. MacDonald, Phys. Rev. B {\bf 80}, 235417 (2009).
\bibitem{Nomura2} K. Nomura, S. Ryu, and D.-H. Lee, Phys. Rev. Lett. {\bf 103}, 216801 (2009).
\bibitem{Hou}C.-Y. Hou, C. Chamon, and C. Mudry, Phys. Rev. B {\bf 81}, 075427 (2010).
\bibitem{Kharitonov} M. Kharitonov, arXiv:1105.6285v1
\bibitem{Gusynin2} V. P. Gusynin, V. A. Miransky, S. G. Sharapov, and I. A. Shovkovy, Phys. Rev. B {\bf 74}, 195429 (2006).
\bibitem{Fuchs} J.-N. Fuchs and P. Lederer, Phys. Rev. Lett. {\bf 98}, 016803 (2007).
\bibitem{Bolotin1} K. I. Bolotin et al., Solid State Commun. 146, 351 (2008).
\bibitem{Das Sarma1} S. Das Sarma and Kun Yang, Solid State Commun. {\bf 149}, 1502 (2009).
\bibitem{Young2011} Very similar results were obtained in graphene samples on hexa boron nitride. See Young et al., unpublished.

\end{thebibliography}
\end{document}